\documentclass[conference,a4paper]{IEEEtran}

\IEEEoverridecommandlockouts

\usepackage{cite}

\usepackage{makeidx,epsfig}
\usepackage{setspace,graphicx,multirow}

\usepackage{amsfonts,latexsym,amssymb,amsthm}

\usepackage{graphicx}
\usepackage{epstopdf}

\usepackage{caption3} 
\DeclareCaptionOption{parskip}[]{} 
\usepackage[small]{caption}

\usepackage{subcaption}

\usepackage{array}

\usepackage[cmex10]{amsmath}

\usepackage{url}

\usepackage{multirow}
\usepackage{hhline}

\usepackage{amsmath}
\usepackage{algorithm}
\usepackage[]{algpseudocode}
\usepackage{varwidth}

\makeatletter
\def\BState{\State\hskip-\ALG@thistlm}
\makeatother

\usepackage{algcompatible}

\usepackage{fixltx2e}

\newcommand*\diff{\mathop{}\!\mathrm{d}}

\usepackage{enumitem}

\makeatletter
  \newcommand\tinyv{\@setfontsize\tinyv{7pt}{9}}
\makeatother

\usepackage{authblk}

\usepackage{textcomp}

\usepackage{dsfont}

\begin{document}
\bibliographystyle{IEEEtran}
\bstctlcite{IEEEexample:BSTcontrol}

\title{Achieving Max-Min Throughput in LoRa Networks}

\author{Jiangbin~Lyu, ~\textit{Member,~IEEE},
		Dan~Yu,
        and~Liqun~Fu, ~\textit{Senior Member,~IEEE}%
\thanks{This work was supported in part by the National Natural Science Foundation 
of China (No. 61801408 and No. 61771017), the Natural Science Foundation of Fujian 
Province (No. 2019J05002) and the Fundamental Research Funds for the Central 
Universities (No. 20720190008).}        
\thanks{The authors are with School of Informatics, and Key Laboratory of Underwater Acoustic Communication and Marine Information Technology, Xiamen University, China 361005 (e-mail: \{ljb, liqun\}@xmu.edu.cn; yudanx@stu.xmu.edu.cn). \textit{Corresponding author: Liqun Fu}.}%
}

\maketitle

\begin{abstract}

With growing popularity, LoRa networks are pivotally enabling Long Range connectivity to low-cost and power-constrained user equipments (UEs). Due to its wide coverage area, a critical issue is to effectively allocate wireless resources to support potentially massive UEs in the cell while resolving the prominent near-far fairness problem for cell-edge UEs,
which is challenging to address due to the lack of tractable analytical model for the LoRa network and its practical requirement for low-complexity and low-overhead design.
To achieve massive connectivity with fairness, we investigate the problem of maximizing the minimum throughput of all UEs in the LoRa network,
by jointly designing high-level policies of spreading factor (SF) allocation, power control, and duty cycle adjustment based only on average channel statistics and spatial UE distribution.
By leveraging on the Poisson rain model along with tailored modifications to our considered LoRa network, we are able to account for channel fading, aggregate interference and accurate packet overlapping, and still obtain a tractable and yet accurate closed-form formula for the packet success probability and hence throughput.
We further propose an \textit{iterative balancing (IB)} method to allocate the SFs in the cell such that the overall max-min throughput can be achieved within the considered time period and cell area.
Numerical results show that the proposed scheme with optimized design greatly alleviates the near-far fairness issue, and significantly improves the cell-edge throughput.%
\end{abstract}

%
\section{Introduction}

%
%
%
%
%
%

The Internet of Things (IoT) has found fast-growing applications over recent years in the civilian domain such as for environmental monitoring, building automation and smart cities, which call for wireless technologies that enable low-cost, large-scale, and ultra-durable connectivity for almost everything.
Low Power Wide Area Network (LPWAN) is one of the IoT paradigms that targets at providing long range wireless connectivity to power-constrained IoT devices in a wide area, which includes pronounced technologies such as Narrow Band (NB)-IoT \cite{NBiotMagazine2016} in the licensed band, and LoRa (Long Range)\cite{LoRaWAN} in the unlicensed band.

This paper focuses on LoRa, one of the most promising LPWAN technologies proposed by Semtech\cite{SemTechLoRaModulationBasics} and further promoted by the LoRa Alliance\cite{LoRaWAN}.
By adaptively trading bit rates for better link budgets, the LoRa physical layer enables flexible long-range communication with low power consumption and low cost design, which is particularly suitable for those user equipments (UEs) that transmit little amount of data over long periods of time, e.g., water and gas meters.
On the other hand, however, due to the wide coverage area, there are potentially massive UEs to be connected by the LoRa Gateway (GW). Worse still, the near-far fairness issue becomes more prominent, as the cell-edge UEs suffer from more severe path-loss and are typically the bottleneck of overall system performance.
To date, how to effectively allocate wireless resources in LoRa networks to support \textit{massive connectivity with fairness in a wide area} remains as a challenging and critical issue.

In particular, the LoRa physical layer adopts the robust Chirp Spread Spectrum (CSS) modulation with different spreading factors (SFs) to accommodate multiple UEs in one channel, whereby the CSS interference comes mostly from the co-SF signals, and has a pseudo-orthogonal characteristic with different-SF signals \cite{SemTechLoRaModulationBasics}.
Moreover, higher SF is associated with lower receiver sensitivity\footnote{Receiver sensitivity is the lowest power level at which the receiver can detect the signal and demodulate data.} (often below the noise floor) thus extending the communication range, at the cost of lower bit rate. 
On top of the LoRa physical layer, LoRa Alliance has defined the higher layers and network architecture termed as LoRaWAN, whereby the medium access control (MAC) layer is essentially an Aloha variant of random access owing to its simplicity. 
In particular, the class A-type devices in LoRaWAN consume the lowest power as they adopt the pure (unslotted) Aloha-like random access with no synchronization or scheduling overhead, and thus are well suited for low duty-cycle devices which are asleep most of the time.
Therefore, we consider class A type in this paper, which is the simplest and also mandatory for all LoRa devices to implement.

Despite the low duty cycle, in very dense deployment scenarios, LoRa networks will still suffer from collisions of concurrent transmissions in the same channel and SF. 
The conventional protocol model\footnote{In the protocol model, two packets are considered both failure if they overlap by any part in time.} for pure Aloha is over simplified which does not account for channel fading, power control, and aggregate interference, and thus cannot characterize the capture effect due to near-far problem in a cell.
A simulation model based on real interference measurements is presented in \cite{LoRaSensors2017}, while a scalability analysis of LoRa networks is performed in \cite{LoRaNS3IoTJ} using a LoRa error model together with the LoRaWAN MAC protocol in the ns-3 simulator.
In \cite{EXPLoRa}, SFs are assigned by equalizing the time-on-air of packets sent by each UE, while in \cite{LoRaPollinICC2017} the transmit power and SFs are assigned to UEs by minimizing the collision probability within the same SF, both of which have not considered the effect of small-scale fading and the fairness in terms of throughput.
The small-scale fading effect has been considered in the classic stochastic geometry method for \textit{slotted} Aloha\cite{SpatialAlohaSlotted}.
In the context of LoRa networks, a mathematical analysis on the uplink coverage is conducted in \cite{LoRaWCL2017Raza} via stochastic geometry, which yet considers only the strongest interferer and ignores the time dependence of (partially) overlapping packets.
Recent works in \cite{2DinterferenceICC2017} and \cite{LoRaCL2018Korean} consider channel fading, aggregate interference, and accurate packet overlapping, and yet it is difficult to obtain the exact distribution of packet success probability analytically.
In summary, a tractable and accurate analytical model is highly needed for the performance evaluation in LoRa networks, which also facilitates further system optimization with low-complexity and low-overhead design.

To achieve massive connectivity with fairness, we study the problem of maximizing the minimum throughput of all UEs in the cell,
by jointly optimizing the SF allocation, power control, and duty cycle adjustment.
Note that achieving the instantaneous max-min throughput usually requires the real-time channel state information (CSI) of all UEs which is prohibitive to collect from low power and duty cycle devices, and worse still, the problem is shown to be a mixed-integer non-linear programming (MINLP) which is computationally costly.
We therefore design high-level control policies based only on average channel statistics and spatial UE distribution, e.g., slow channel inversion power control based on average channel strength.
Moreover, for analytical tractability, we leverage another thread of stochastic geometry for \textit{non-slotted} Aloha, i.e., the Poisson rain model\cite{NonSlottedINFOCOM2010}, which caters for channel fading, aggregate interference, and accurate packet overlapping. 
With tailored modifications to our considered LoRa network, we are able to obtain a tractable and yet accurate closed-form formula for the packet success probability and hence throughput, based on which the optimal duty cycle can be derived.
We further propose an \textit{iterative balancing (IB)} method to allocate the SFs in the cell such that the overall max-min throughput can be achieved within the considered time period and cell area, even when the duty cycle is pre-determined and non-adjustable.
Numerical results show that the proposed scheme with optimized design greatly alleviates the near-far fairness issue in a large service area, and significantly improves the cell-edge throughput. Moreover, the total spatial throughput has also been improved significantly, compared with benchmark schemes with fixed transmit power and duty cycle. 
Finally, besides system optimization, the proposed throughput formula is itself useful for performance evaluation, while the proposed analytical framework can be extended to other general settings in LoRa networks.

\textit{Notations}: $\mathbb{R}$ denotes the set of real numbers; $\mathbb{Z}$ denotes the set of integer numbers; $\mathbb{P}\{\cdot\}$ denotes the probability of an event; $\mathbb{E}\{\cdot\}$ denotes the expectation of a random variable (RV); $\stackrel{\textrm{dist.}}{=}$ denotes equal in distribution; $\exp\{\cdot\}$ denotes the exponential function; $\|\cdot\|$ denotes the Euclidean norm; $|\cdot|$ takes the cardinality of a set; $\setminus\cdot$ denotes the set minus operation; $\bigcup$ denotes the set union; $\bigcap$ denotes the set intersection; and $\emptyset$ denotes the empty set.

\section{System Model}\label{SectionModel}

\begin{figure}
\centering
   \includegraphics[width=0.5\linewidth,  trim=0 0 0 0,clip]{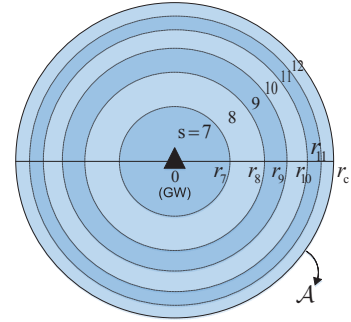}
\caption{Single-gateway LoRa networks.\vspace{-4ex}}\label{LoRa}
\end{figure}

Consider uplink\footnote{Uplink communication from LoRa end devices to the network server is typically uncoordinated and presents as the performance bottleneck.} communication from distributed UEs to a single GW with height $H_G$ meters (m) at the origin and a disk cell area $\mathcal{A}\subset\mathbb{R}^2$ of radius $r_c$ m, as shown in Fig. \ref{LoRa}.
In the considered time period, denote $\mathcal{K}=\{1,2,\cdots,K\}$ as the set of active UEs\footnote{We neglect ``active" in the rest of the paper for simplicity.} in the cell which have packets to transmit, with 2D locations $\bold w_k\triangleq (x_k,y_k),x_k,y_k\in\mathbb{R}, k\in\mathcal{K}$ on the ground, which are assumed to
follow a homogeneous Poisson point process (HPPP) $\Phi\subset \mathcal{A}$ with density $\lambda$ /m$^2$.
Then $K$ is a Poisson RV with mean $\bar{K}\triangleq \lambda\pi r_c^2$.

LoRa is adopted as the physical layer transmission technology for UE-GW communications.
For the purpose of exposition, consider a single channel with bandwidth $B$ Hz, and pseudo-orthogonal SFs $s\in\mathcal{S}\triangleq\{7,\cdots,12\}$.
For a given $s$, the transmission rate in bits/second (bps) is given by\cite{SemTechLoRaModulationBasics}
\begin{equation}
R_s=\frac{s}{2^s}BC,
\end{equation}
where $C$ is the code rate.
For simplicity, assume that the packets are of equal length $L_s$ bits for a given $s$, which corresponds to a packet duration of $T_s\triangleq L_s/R_s$.

Denote $P_k$ as the transmit power of UE $k$ in Watt (W), capped by the maximum allowed value $P_\textrm{max}$.
Assume that the GW and UEs are each equipped with a single omnidirectional antenna with unit gain. 
We assume a fading channel between the GW and UEs, which consists of distance-dependent path-loss with path-loss exponent $n\geq 2$ and an additional random term accounting for small-scale fading.
Therefore, the channel power gain from the GW to UE $k$ can be modelled as
$g_k=\bar g_k\zeta_k$, where $\bar g_k\triangleq\alpha_0 (H_G^2+d_k^2)^{-n/2}$ is the average channel power gain, with $\alpha_0=(\frac{4\pi f_c}{c})^{-2}$ denoting the average channel power gain at a reference distance of 1 m, $f_c$ denoting the carrier frequency, $c$ denoting the speed of light, and $d_k$ denoting the horizontal distance between GW 0 and UE $k$;
 and $\zeta_k\stackrel{\textrm{dist.}}{=}\zeta\sim \textrm{Exp}(1/\mu)$ is an independent and identically distributed (i.i.d.) exponential RV with mean $1/\mu$ accounting for the small-scale Rayleigh fading.\footnote{The proposed analytical method in this paper can be readily extended to account for other fading channels.}

Consider a typical UE 0 with SF $s$.
Assume that the receiver noise at the GW is additive white Gaussian noise (AWGN) with zero mean and power $\sigma^2$ (W).
In the case without interference, the received signal-to-noise ratio (SNR) is given by
\begin{equation}
\eta_{s,0}\triangleq  P_0 g_0/\sigma^2= P_0\bar g_0\zeta_0/\sigma^2,
\end{equation}
which needs to be not smaller than a certain threshold $\bar\eta_{s}$ in order for the packet to be successfully decoded,
where $\bar\eta_{s}$ is typically lower for higher SF (see Table I in \cite{LoRaWCL2017Raza}).

In the case when multiple UEs transmit concurrently on the same channel and SF, the co-SF interference could fail the overlapping packets.
Consider pure (unslotted) Aloha as the MAC-layer multiple access method, which corresponds to Class A in LoRaWAN.
Denote $\mathcal{K}_s\subseteq \mathcal{K}$ as the set of UEs with SF $s$, i.e.,
\begin{equation}
\mathcal{K}_s=\{k|s_k=s, k\in\mathcal{K}\},
\end{equation}
where $s_k$ is the SF of UE $k$.
Denote $K_s\triangleq |\mathcal{K}_s|$ as the number of UEs with SF $s$.
Denote the duty cycle, i.e., fraction of time that an active UE is transmitting packets, as $\delta_k\in [0,1]$ for UE $k$, which is subject to a certain limit $\Delta_\textrm{max}$ (e.g., 1\%).
For simplicity, assume that the UEs in $\mathcal{K}_s$ adopt the same duty cycle $\delta_k=\Delta_s, \forall k\in \mathcal{K}_s$.
The number of transmission initiations per UE per unit time is then given by
\begin{equation}\label{rho_s}
\rho_s\triangleq \frac{1}{T_s/\Delta_s-T_s}=\frac{\Delta_s}{(1-\Delta_s)T_s},
\end{equation}
which corresponds to the time frequency of channel access.

The aggregate co-SF interference is modeled in the following.
Consider the typical UE 0 with SF $s$ which transmits a reference packet during the time interval $[0,T_s]$.
The average number of co-SF packets initiated during this packet interval is given by $K_s \rho_s T_s$.
At a given time instant $t\in [0,T_s]$, the transmitting UEs with SF $s$ make up a set $\mathcal{K}_s(t)\subseteq\mathcal{K}_s$, which cause the total received interference power at the GW as 
\begin{equation}\label{Ist}
I_s(t)\triangleq \sum_{k\in\mathcal{K}_s(t)\setminus\{0\}} P_k g_k=\sum_{k\in\mathcal{K}_s(t)\setminus\{0\}} P_k \bar g_k \zeta_k.
\end{equation}
The aggregate interference power at the GW averaged over one packet duration $T_s$ is then given by
\begin{equation}\label{barIs}
\bar I_s\triangleq \frac{1}{T_s}\int_{0}^{T_s} I_s(t) \diff t.
\end{equation}

Under the non-slotted Aloha model with average interference constraint\cite{NonSlottedINFOCOM2010}, 
the signal-to-interference ratio (SIR) $\gamma_{s,0}$ of the reference packet is given by
\begin{equation}\label{SINR0}
\gamma_{s,0}\triangleq P_0\bar g_0\zeta_0/\bar I_s,
\end{equation}
which corresponds to a situation where some coding with repetition and interleaving of bits on the whole packet duration is used; and 
the reference packet can be successfully decoded if its SIR is not smaller than a certain threshold $\bar\gamma_s$ (see Table 1 in \cite{LoRaThreshold2015}).

In summary, the success probability of a reference packet sent by the typical UE 0 with SF $s$ is thus given by
\begin{align}
\textrm{P}_{s,0}^{\textrm{suc}}&\triangleq\mathbb{P}\big\{\{\eta_{s,0}\geq\bar\eta_s\} \&\{\gamma_{s,0}\geq\bar\gamma_s\}\big\}\label{PsusTrue}\\
&=\mathbb{P}\big\{\{P_0\bar g_0\zeta_0/\sigma^2\geq\bar\eta_s\} \&\{P_0\bar g_0\zeta_0/\bar I_s\geq\bar\gamma_s\}\big\}\notag\\
&\stackrel{(a)}{\geq}\mathbb{P}\big\{\zeta_0\geq\bar\eta_s\sigma^2/(P_0\bar g_0)\big\} \mathbb{P}\big\{\zeta_0\geq\bar\gamma_s\bar I_s/(P_0\bar g_0)\big\}\notag\\
&\stackrel{(b)}{=}e^{-\mu\bar\eta_s\sigma^2/(P_0\bar g_0)}\mathbb{E}_{\bar I_s}\big\{e^{-\mu\bar\gamma_s\bar I_s/(P_0\bar g_0)}\big\},\label{Psuc0}
\end{align}
where the inequality $(a)$ holds since the two events on $\zeta_0$ are positively correlated\footnote{We have $\mathbb{P}\big\{\{\zeta_0\geq a\}\&\{\zeta_0\geq b\}\big\}=\mathbb{P}\big\{\zeta_0\geq a\big\}\mathbb{P}\big\{\zeta_0\geq b|\zeta_0\geq a\big\}$, where $\mathbb{P}\big\{\zeta_0\geq b|\zeta_0\geq a\big\}\geq \mathbb{P}\big\{\zeta_0\geq b\big\}$ since the event $\zeta_0\geq a$ implies that $\zeta_0\geq b$ is more likely to happen.}; $(b)$ is due to $\zeta_0$ with exponential distribution; and $\mathbb{E}_{\bar I_s}\big\{e^{-\mu\bar\gamma_s\bar I_s/(P_0\bar g_0)}\big\}\triangleq \mathcal{L}_{\bar I_s}\big(\mu\bar\gamma_s/(P_0\bar g_0)\big)$ where $\mathcal{L}_{\bar I_s}(\cdot)$ is the Laplace transform\footnote{The Laplace transform of an RV $X$ is defined as $\mathcal{L}_{X}(z)\triangleq \mathbb{E}_{X}\big\{e^{-zX}\big\}$.} of the RV $\bar I_s$.
Note that the inequality $(a)$ is tight when $\bar\gamma_s\bar I_s\gg \bar\eta_s\sigma^2$, which is typically the case for interference-limited scenario, especially thanks to the low value of $\bar\eta_s$ (e.g., -6 dB to -20 dB) associated with the low sensitivity of LoRa receiver.

As a result, we need to characterize the Laplace transform of the average interference $\bar I_s$ in order to obtain $\textrm{P}_{s,0}^{\textrm{suc}}$, and further obtain the throughput of the typical UE 0 in bps given by
\begin{equation}
\theta_{s,0}\triangleq R_s\Delta_s\textrm{P}_{s,0}^{\textrm{suc}}.
\end{equation}
The throughput of other UE $k$ with SF $s$, denoted as $\theta_{s,k}$, can be obtained similar to $\theta_{s,0}$.

\section{Problem Formulation}

In this section, we formulate the optimization problem to maximize the minimum throughput $\theta$ of all UEs by jointly optimizing the transmit power $P_k$, SF $s_k$, and duty cycle $\Delta_s$ of all UEs $k\in\mathcal{K}$, under given UE density $\lambda$ and cell radius $r_c$.
The problem can be formulated as
\begin{align}
\mathrm{(P1)}: \underset{
\begin{subarray}{c}
  \theta,\Delta_s, s\in\mathcal{S}\\
  P_k, s_k, k\in\mathcal{K}
  \end{subarray}
}{\max}& \quad\theta \notag\\
             \text{s.t.}\quad&\theta_{s,k}\geq \theta,\quad \forall k\in\mathcal{K}_s, s\in\mathcal{S},\label{ConstraintTheta}\\ 
             &0\leq \Delta_s\leq \Delta_\textrm{max},\quad s\in\mathcal{S},\label{ConstraintDelta}\\            
             &0\leq P_k\leq P_\textrm{max},\quad k\in\mathcal{K},\label{ConstraintP}\\ 
             &s_k\in\mathcal{S},\quad k\in\mathcal{K}.\label{ConstraintS}       
\end{align}

The problem (P1) is a MINLP due to the discrete SF allocation and the non-linear constraint \eqref{ConstraintTheta}.
In general, it requires $O(|\mathcal{S}|^{K})$ complexity to search for the optimal solution, which is prohibitive for the scenario with massive IoT devices.

In addition to the complexity, solving problem (P1) also faces practical challenges.
Due to the low energy consumption requirement of IoT devices, the UEs typical have low wake-up frequency and limited active time upon wake-up. 
Therefore, the set of active UEs may be constantly changing over time from the pool of massive IoT devices.
As a result, it would be difficult for the GW to obtain the instantaneous CSI of all UEs at all time which is required to solve (P1) optimally.
Moreover, the GW has limited downlink capacity to send control or feedback information to UEs.
Therefore, even if the centralized problem (P1) can be solved instantaneously for every given snapshot of active UEs, it would create a huge overhead for GW-UE handshaking and conveying the optimal solutions to each individual UE.

In order to resolve the above challenges, we propose to design high-level \textit{policies} to optimize the overall system performance in a larger time scale.
Specifically, instead of requiring detailed information about UE locations and instantaneous CSI at a given time snapshot, we assume only knowledge about the active UE distribution and density over a certain period of time as well as the channel statistics such as fading distribution and path-loss exponent.
We seek to maximize the minimum throughput averaged over time, by jointly designing the policies of power control, SF allocation and duty cycle adjustment.

\section{Proposed Scheme}

In general, higher SF is associated with lower SNR threshold $\bar\eta_s$ at the cost of lower data rate\cite{SemTechLoRaModulationBasics}, which helps to extend the communication range from the GW in the absence of interference.
Therefore, it is natural to assign higher SF to UEs at a longer distance from the GW, which have larger path-loss and hence lower average received power at the GW under the same transmit power.
We thus adopt the distance-based SF allocation policy without requiring real-time CSI.

Specifically, for the considered cell area of radius $r_c$, it is partitioned into $|\mathcal{S}|=6$ zones with the delimiting distance threshold $r_s, s=7,\cdots,11$. By default, denote $r_6=0$ and $r_{12}=r_c$. The UEs in each zone $\mathcal{A}_s$ within distance $r_{s}$ to $r_{s-1}$ are allocated with the SF $s$, respectively.
For example, the UEs within distance $r_7$ from GW 0 are allocated with SF 7, the UEs within distance $r_8$ to $r_7$ are allocated with SF 8 and so on, as shown in Fig. \ref{LoRa}.
Denote $\bold r\triangleq (r_7,\cdots,r_{11})$ as the vector of partitioning distance threshold, which is the optimization variable to determine the SF allocation.
For HPPP distributed UEs, the number of UEs $K_s$ allocated to SF $s$ is a Poisson RV with mean $\lambda A_s$, where $A_s=|\mathcal{A}_s|$ denotes the area associated with SF $s$ and is given by
\begin{equation}\label{Area}
A_s=\pi (r_s^2-r_{s-1}^2), s=7,\cdots,12.
\end{equation}

When the partitioning distance threshold $\bold r$ is given, the power control problem is also simplified.
If the typical UE is at the zone edge with distance $r_s$ from the GW, its packet success probability follows from \eqref{Psuc0}, and thus is more likely to be in outage due to larger path-loss than the inner UEs with the same SF $s$, under the same transmit power.
In order to achieve max-min throughput for the UEs $k\in\mathcal{K}_s$ without real-time CSI, we propose the ``slow" channel inversion power control based on the average channel power gain $\bar g_k$, such that the average received power at the GW is the same\footnote{For simplicity, continuous power control is considered here while the obtained results can be quantized into discrete power levels for implementation.} for all UEs with SF $s$, denoted by $\bar Q_{s}$.
Specifically, the transmit power of each UE $k\in\mathcal{K}_s$ is given by
\begin{equation}\label{PowerControlk}
P_k=\bar Q_{s}/\bar g_k=\bar Q_{s}(H_G^2+d_k^2)^{n/2}/\alpha_0, \forall k\in\mathcal{K}_s,
\end{equation}
where the transmit power $P_k$ is inversely proportional to $\bar g_k$, and thus we have $\bar Q_{s}=P_k\bar g_k, \forall k\in\mathcal{K}_s$, including the typical UE 0.
The power control in \eqref{PowerControlk} can also be written in the form of distance-based policy as follows:
\begin{equation}
P(s,r)\triangleq \bar Q_{s}(H_G^2+r^2)^{n/2}/\alpha_0,
\end{equation}
where $P(s,r)$ denotes the transmit power of the UE with SF $s$ at distance $r$ from the GW.
In particular, for the UE at the zone edge with transmit power $P_s^\textrm{edge}\triangleq P(s,r_s)$, we have 
\begin{equation}\label{Qs}
\bar Q_{s}\triangleq P_s^\textrm{edge}\alpha_0(H_G^2+r_s^2)^{-n/2},
\end{equation}
and hence
\begin{equation}
P(s,r)=P_s^\textrm{edge}\bigg(\frac{H_G^2+r^2}{H_G^2+r_s^2}\bigg)^{n/2}.
\end{equation}
In other words, the transmit power of inner UEs with SF $s$ and $r<r_s$ is \textit{reduced} so that their average received power at the GW equals to that of the UE at the zone edge, which achieves both fairness and power savings.

Under the above UE partitioning and power control, we derive the packet success probability in the following. 
The average interference $\bar I_s$ is given in \eqref{barIs}, and the instantaneous interference $I_s(t)$ in \eqref{Ist} is reduced to
\begin{equation}\label{Istr}
I_s(t)= \sum_{k\in\mathcal{K}_s(t)\setminus\{0\}} \bar Q_s \zeta_k.
\end{equation}
Note that the UEs $k\in\mathcal{K}_s$ reside in the ring region $\mathcal{A}_s$, whose locations $\bold w_k$ and packet initiation time instant $t_k$ are both random, rendering the set $\mathcal{K}_s(t)$ difficult to model and analyze.

In this paper, we adopt the Poisson rain model for the UEs $k\in\mathcal{K}_s(t)$, which forms a space-time HPPP $\Phi_s\triangleq\{(\bold w_k, t_k), k\in\mathcal{K}_s\}$. We may think of $\bold w_k$ ``born" at time instant $t_k$ transmitting a packet during time interval $[t_k,t_k+T_s)$ and ``disappearing" immediately after. The HPPP $\Phi_s$ has a density $\lambda \rho_s$, which corresponds to the space-time frequency of channel access.
In the sequel, we derive the Laplace transform of the interference $\bar I_s$ based on the formula for the Laplace functional of the HPPP:
\newtheorem{lemma}{Lemma}
\begin{lemma}[Fact A.3 in \cite{NonSlottedINFOCOM2010}]\label{lem1}
Consider a generic shot-noise $J\triangleq\sum_{Y_k\in\Pi}f(\zeta_k,Y_k)$ generated by some HPPP $\Pi$ with density $\Lambda$, response function $f(\cdot,\cdot)$ and i.i.d. marks $\zeta_k$ distributed as a generic RV $\zeta$. Then the Laplace transform of $J$ is given by
\begin{equation}\label{ShotNoiseLT}
\mathcal{L}_J(z)=\exp\bigg\{-\Lambda \int\big(1-\mathbb{E}_\zeta\{e^{-zf(\zeta,y)}\}\big)\diff y\bigg\},
\end{equation}
where the integral is evaluated over the whole state space on which $\Pi$ lives.
\end{lemma}

Based on Lemma \ref{lem1}, we derive the Laplace transform of the average interference $\bar I_s$ in our setting with power control and bounded UE distribution field.
The results are summarized in the following proposition:
\newtheorem{prop}{Proposition}
\begin{prop}\label{prop1}
The Laplace transform of the average interference $\bar I_s$ defined by \eqref{barIs} and \eqref{Istr} under the Poisson rain model is given by
\begin{equation}
\mathcal{L}_{\bar I_s}(z)=\exp\bigg\{\frac{-2\lambda A_s\Delta_s}{1-\Delta_s} \bigg(1+\frac{\mu}{\bar Q_s z}\ln\frac{\mu}{\mu+\bar Q_s z}\bigg)\bigg\}.
\end{equation}
\end{prop}

\textit{Proof:} Please refer to Appendix \ref{AppendixProp1}.$\blacksquare$

For the UEs $k\in\mathcal{K}_s$ with i.i.d. fading $\zeta_k\stackrel{\textrm{dist.}}{=}\zeta$ and under the above power control, their SNR and SIR are both equal in distribution, respectively, and hence they have equal packet success probability which can be obtained based on \eqref{Psuc0} and Proposition \ref{prop1} as follows:
\begin{align}\label{PsusFinal}
\textrm{P}_{s,0}^{\textrm{suc}}&=e^{-\mu\bar\eta_s\sigma^2/\bar Q_{s}}\mathbb{E}_{\bar I_s}\big\{e^{-\mu\bar\gamma_s\bar I_s/\bar Q_{s}}\big\}\notag\\
&=\exp\bigg\{\frac{-\sigma^2\mu\bar\eta_{s}}{\bar Q_{s}}-\frac{2\lambda A_s C_s\Delta_s}{1-\Delta_s}\bigg\},
\end{align}
where $C_s\triangleq \big(1+\frac{1}{\bar\gamma_s}\ln\frac{1}{1+\bar\gamma_s}\big)$ is a constant.
Therefore, the common throughput of UEs with SF $s$ is given by
\begin{equation}\label{ThetasFinal}
\bar \theta_s\triangleq R_s\Delta_s\textrm{P}_{s,0}^{\textrm{suc}}=R_s\Delta_s\exp\bigg\{\frac{-\sigma^2\mu\bar\eta_{s}}{\bar Q_{s}}-\frac{2\lambda A_s C_s\Delta_s}{1-\Delta_s}\bigg\}.
\end{equation}

The tractable closed-form formula in \eqref{ThetasFinal} reveals insights about the effects of key system parameters.
First, $\bar \theta_s$ is monotonically increasing with the equalized received power $\bar Q_s$ in \eqref{Qs}, which suggests $P_s^{\textrm{edge}}=P_\textrm{max}$ to be adopted by the zone-edge UE. 
Second, $\bar \theta_s$ is monotonically decreasing with the zone-edge distance $r_s$, which affects not only the average received power $\bar Q_s$ at the zone edge, but also the zone area $A_s$ (and hence the number of UEs in the zone).
Third, it can be verified that $\bar \theta_s$ first increases and then decreases with $\Delta_s$, which facilitates efficient solutions to achieve maximum throughput for those UEs with SF $s$. Also note that $\bar \theta_s$ goes to 0 as $\Delta_s\rightarrow 1$, which is consistent with the practical scenario where all UEs keep transmitting all the time.
Finally, besides for system optimization, the throughput formula \eqref{ThetasFinal} is itself useful as a tractable analytical result for performance evaluation in LoRa networks, while its deriving method is extendable to other general settings, e.g., multiple-GW scenario.

To this end, the problem (P1) can be re-cast into the following problem:
\begin{align}
\mathrm{(P2)}: \underset{
\begin{subarray}{c}
  \bar\theta, \bold r\\
  \Delta_s, s\in\mathcal{S}
  \end{subarray}
}{\max}& \quad\bar\theta \notag\\
             \text{s.t.}\quad&\bar\theta_s\geq \bar\theta,\quad s\in\mathcal{S},\label{ConstraintTheta1}\\ 
             &0\leq \Delta_s\leq \Delta_\textrm{max},\quad s\in\mathcal{S},\label{ConstraintDelta1}      
\end{align}
where the optimal solution to (P2) is denoted as $(\bold r^*;\Delta_s^{*}, s\in\mathcal{S})$, and the corresponding max-min throughput denoted as $\bar\theta^{*}$.
In order to solve (P2), we first obtain the optimal duty cycle under given $\bold r$ for each SF $s$ as follows:
\begin{equation}\label{DutyCycleOpt}\small
\Delta_s^*(\bold r)=\min\bigg\{\Delta_\textrm{max},\frac{1}{1+\lambda A_s C_s+\sqrt{\lambda A_s C_s(2+\lambda A_s C_s)}}\bigg\},
\end{equation}
which is obtained by finding the root of the first-order derivative on $\Delta_s$ in \eqref{ThetasFinal} along with the constraint \eqref{ConstraintDelta1}.
It can be further verified that the corresponding throughput $\bar\theta_s^*(\bold r)$ is decreasing with the zone area $A_s$, which itself is decreasing with $r_{s-1}$ but increasing with $r_s$ as given in \eqref{Area}.

Based on such monotonicity, we propose the \textit{Iterative Balancing (IB)} method to find the optimal partitioning threshold $\bold r^*$ to achieve the overall max-min throughput $\bar\theta^{*}$.
Specifically, in each iteration, we find two neighboring zones $\mathcal{A}_s$ and $\mathcal{A}_{s+1}$ which have the largest throughput gap $\vartheta_{\textrm{max}}$, and then tune $r_s$ with others in $\bold r$ fixed, such that the throughput gap is eliminated or reduced to the best extend.
The iterations continue until $\vartheta_{\textrm{max}}$ is less than a certain threshold $\epsilon$, or until none of the throughput gaps, in descending order, can be further reduced.
Since the throughput gaps are bounded below by 0, and the larger gap is decreasing in each iteration, convergence is guaranteed at which the (lexicographical) max-min throughput is achieved.
Due to space limit, the detailed algorithm pseudo code can be found in \cite{LoRaMaxMinArXiv}.
Finally, note that for the case where the duty cycle is pre-determined and cannot be adjusted, the proposed power control and IB method can still be applied to achieve the corresponding max-min throughput given the duty cycle.

\section{Numerical Results}



In this section, we first verify the accuracy of the analytical formula in \eqref{ThetasFinal} by comparing with Monte Carlo (MC) simulation results, which is then applied to obtain the max-min throughput solution to (P2) under different cell radius $r_c$.
The following parameters are used if not mentioned otherwise: $H_G=25$ m, $\lambda=700$/km$^2$, $B=125$ kHz, $C=4/5$, $L_s=25$ Bytes, $P_{\textrm{max}}=14$ dBm, $n=3.5$, $f_c=868$ MHz, $c=3\times 10^8$ m/s, $\mu=1$, $\sigma^2=-117$ dBm, $\Delta_{\textrm{max}}=1\%$, $\epsilon=0.02$ bps, $\bar\gamma_s=6$ dB for $s\in\mathcal{S}$, and $\bar\eta_s$ is given in Table \ref{table1}.

\begin{table}[t]\scriptsize
\caption{LoRa parameters}
\addtolength{\tabcolsep}{-4pt}
\renewcommand{\arraystretch}{1.1}
\centering
\begin{tabular}{|c|c|c|c|c|}
\hline
SF&BitRate$R_s$(bps)&SNR threshold $\bar\eta_s$(dB)&Max Range(m)&EqualArea Range(m)\\ \hline
7 & 5469 & -6 & 1053 & 408 \\ \hline
8 & 3125 & -9 & 1283 & 577\\ \hline
9 & 1758 & -12 & 1563 & 707\\ \hline
10 & 977 & -15 & 1904 & 816\\ \hline
11 & 537 & -17.5 & 2244 & 913\\ \hline
12 & 293 & -20 & 2645 & 1000\\ \hline
\end{tabular}
\centering
\label{table1}
\end{table}



In the MC simulations, 
we first generate the location database for the set $\mathcal{K}_{\textrm{all}}$ of all UEs (including both active and inactive) in the considered area, which is a random realization of an HPPP with larger density $\lambda_{\textrm{all}}$ (e.g., $\lambda_{\textrm{all}}=2\lambda$).
Then the set $\mathcal{K}$ of active UEs in the considered time period is randomly and independently drawn from $\mathcal{K}_{\textrm{all}}$.
The packet success probability is obtained by averaging over $N$ (e.g., $N=10^6$) realizations of $\mathcal{K}$, where for each realization we simulate the random packet generation and verify both the SNR and SIR conditions for the reference packet according to \eqref{PsusTrue}.

\begin{figure}[t]
  \centering
  \vspace{-1em}
  \includegraphics[width=1\linewidth]{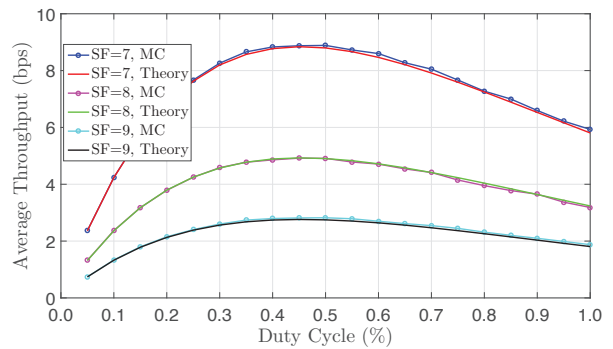}
  \caption{MC simulation verification for the throughput formula \eqref{ThetasFinal}.}\label{fig1}
   \vspace{-0.5em}
\end{figure}

\begin{figure}[t]
  \centering
  \includegraphics[width=1.0\linewidth]{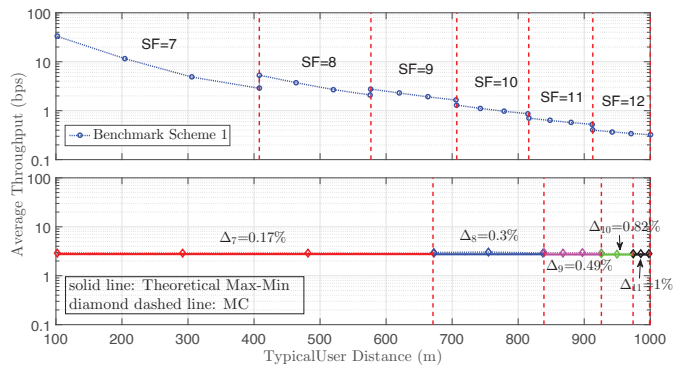}
  \caption{Throughput distribution in a cell with radius 1 km.}\label{fig2}
   \vspace{-1em}
\end{figure}

In the first set of simulations, for the purpose of illustration, we consider a cell with radius $r_c=1$ km which is partitioned into six equal-area ring regions (similar to Fig. \ref{LoRa}) each allocated with SF 7 to 12, respectively.
Under the slow channel inversion power control, we obtain the average (common) throughput of the UEs in each region for a given duty cycle, and plot the results in Fig. \ref{fig1}\footnote{SFs 10, 11 and 12 have similar results which are omitted for brevity.}, for both the MC method and our proposed throughput formula \eqref{ThetasFinal}. 
It can be seen that the proposed formula matches quite well with the MC simulation results.
Therefore, in the following, we apply \eqref{ThetasFinal} to find the optimal operating point for achieving the overall max-min throughput in (P2) under different target service area.

In the second set of simulations, we first simulate a benchmark scheme 1 using the MC method with fixed transmit power $P_k=P_{\textrm{max}}$ and fixed duty cycle $\delta_k=1\%$ for all UEs $k\in\mathcal{K}$, under the equal-area cell partitioning where the partitioning distance threshold $\bold r$ is given in the last column of Table \ref{table1}.
The results are plotted in the upper half of Fig. \ref{fig2}. It can be seen that the average throughput of the typical UE with a certain SF decreases with its distance from the GW, resulting in a prominent near-far fairness issue. Worse still, the UEs close to the cell edge (with SF 11 or 12) suffer from a poor throughput due to lower bit rate.

In comparison, in the lower half of Fig. \ref{fig2}, we plot the results obtained by our proposed scheme for solving (P2), which matches well with the MC simulations under the obtained operating point.
It can be seen that the achieved max-min throughput is around 3 bps per UE, where more UEs tend to be allocated with lower SF to enjoy higher bit rate. 
In particular, the optimal duty cycle of SF 11 is capped by 1\%, while SF 12 is not used in this case due to its low bit rate.
As a result, the near-far fairness issue is greatly alleviated, and the throughput of cell-edge UEs greatly improved. 
On the other hand, define the spatial throughput of all UEs $k\in\mathcal{K}$ in bps/m$^2$ as the ratio of the total throughput versus the considered cell area, i.e.,
\begin{equation}
\Theta\triangleq \frac{\sum_{s\in\mathcal{S}}\sum_{k\in\mathcal{K}_s} \theta_{s,k}}{\pi r_c^2}.
\end{equation}
Compared with the benchmark scheme 1, the overall spatial throughput has also been improved from 142 to 1000 bps/km$^2$.

\begin{figure}[t]
  \centering
  \includegraphics[width=1.02\linewidth]{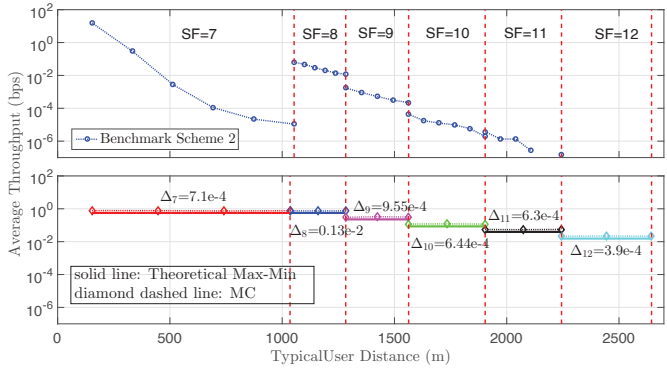}
  \caption{Throughput distribution in a cell with radius 2645 m.}\label{fig3}
   \vspace{-1em}
\end{figure}

In the last set of simulations, we consider a larger cell with radius $r_c=2645$ m, which is the maximum range reachable by SF 12 under path-loss only.
A benchmark scheme 2 is simulated using the MC method with fixed transmit power and duty cycle as in benchmark scheme 1, while the partitioning distance threshold $r_s$ is set as the maximum range reachable by SF $s$ under path-loss only, as in the fourth column of Table \ref{table1}.
The results are plotted in the upper half of Fig. \ref{fig3}. Due to longer range and larger service area, it can be seen that the near-far fairness issue becomes more severe, and the throughput of cell-edge UEs becomes even poorer. 
In comparison, we also plot the results obtained by our proposed scheme in the lower half of Fig. \ref{fig3}, which greatly alleviates the near-far fairness issue and improves the cell-edge throughput. 
Compared with the benchmark scheme 2, the overall spatial throughput has also been improved from 10.3 bps/km$^2$ to 88.2 bps/km$^2$.
Finally, some deployment/design guidelines are suggested for a large service area. The properly-optimized LoRa network is able to serve the cell-edge UEs but typically at a low throughput, and it is beneficial to allow the UEs to adjust their duty cycle so as to alleviate the collisions from massive co-SF devices.

\section{Conclusions}

To achieve massive connectivity with fairness in LoRa networks, this paper investigates the max-min throughput problem by jointly designing high-level policies of SF allocation, power control, and duty cycle adjustment based only on average channel statistics and spatial UE distribution.
By leveraging on the Poisson rain model along with tailored modifications to our considered LoRa network, we are able to obtain a tractable and yet accurate closed-form formula for the packet success probability and hence throughput.
We further propose an IB method to allocate the SFs in the cell such that the overall max-min throughput can be achieved.
Numerical results show that the proposed scheme with optimized design greatly alleviates the near-far fairness issue, and significantly improves the cell-edge throughput.
Moreover, the overall spatial throughput has also been improved significantly compared with benchmark schemes with fixed transmit power and duty cycle. 
Future work could extend the current scheme to the multiple-GW scenario to further improve the throughput and coverage range.
\appendices
\section{}\label{AppendixProp1}
The average interference $\bar I_s$ follows from \eqref{barIs} and \eqref{Istr}:
\begin{equation}\label{barIsShot}\small
\bar I_s= \frac{1}{T_s}\int_{0}^{T_s} \sum_{k\in\mathcal{K}_s(t)\setminus\{0\}} \bar Q_s \zeta_k \diff t\stackrel{\textrm{dist.}}{=}\sum_{(\bold w_k, t_k)\in\Phi_s}\zeta_k h(t_k)\bar Q_s,
\end{equation}
where the second equality is due to swapping of integration and summation, and the fact that the HPPP remains equal in distribution after removing one point; and 
\begin{equation}\small
h(t_k)\triangleq\frac{1}{T_s}\int_{0}^{T_s} \bold{1}(t_k\leq t<t_k+T_s) \diff t=\frac{\max\big(T_s-|t_k|,0\big)}{T_s},
\end{equation}
which represents the time ratio of UE $k$'s packet overlapping with the reference packet in the time interval $[0, T_s]$,
where $\bold{1}(\cdot)$ is the indicator function.

Note that \eqref{barIsShot} matches with the shot-noise definition with the response function $\zeta_k h(t_k)\bar Q_s$, and thus it follows from Lemma \ref{lem1} that
\begin{align}
\mathcal{L}_{\bar I_s}(z)&=e^{-\lambda\rho_s \int_{-\infty}^{\infty}\int_{0}^{2\pi}\int_{r_{s-1}}^{r_s}\big(1-\mathbb{E}_\zeta\{e^{-z\zeta h(t)\bar Q_s}\}\big)r\diff r\diff \phi \diff t}\notag\\
&=e^{-\lambda\rho_s A_s \int_{-\infty}^{\infty}\big(1-\mathbb{E}_\zeta\{e^{-z\zeta h(t)\bar Q_s}\}\big) \diff t}\notag\\
&\stackrel{(a)}{=}e^{-\lambda\rho_s A_s \int_{-\infty}^{\infty}\big(1-\frac{\mu}{\mu+zh(t)\bar Q_s}\big) \diff t}\notag\\
&=e^{-2\lambda\rho_s A_s T_s\big(1+\frac{\mu}{\bar Q_s z}\ln\frac{\mu}{\mu+\bar Q_s z}\big)},\label{LaplaceDerive}
\end{align}
where $A_s$ is given in \eqref{Area}; and $(a)$ is due to the Laplace transform of exponentially distributed $\zeta$ with mean $1/\mu$, which is given by $\mathcal{L}_{\zeta}(z')\triangleq \frac{\mu}{\mu+z'}$. Therefore, Proposition \ref{prop1} follows by substituting \eqref{rho_s} into \eqref{LaplaceDerive}.$\blacksquare$

\bibliography{IEEEabrv,BibDIRP}

\newpage

\end{document}